\documentclass[12pt]{article}
\usepackage{epsfig}
\begin{document}
\baselineskip 18pt
\newcommand{\Tr}{\mbox{Tr\,}}
\newcommand{\Dirac}{/\!\!\!\!D}
\newcommand{\beq}{\begin{equation}}
\newcommand{\eeq}[1]{\label{#1}\end{equation}}
\newcommand{\bea}{\begin{eqnarray}}
\newcommand{\eea}[1]{\label{#1}\end{eqnarray}}
\renewcommand{\Re}{\mbox{Re}\,}
\renewcommand{\Im}{\mbox{Im}\,}
\begin{titlepage}
\hfill  CERN-TH/98-323 IFUM-633-FT IMPERIAL/TP/98-99/4
NYU-TH/98/10/03\vskip .01in \hfill hep-th/9810126
\begin{center}
\hfill
\vskip .4in
{\large\bf Novel Local CFT and Exact Results on Perturbations of N=4 Super
Yang Mills from AdS Dynamics}
\end{center}
\vskip .4in
\begin{center}
{\large L. Girardello$^{a,b}$, M. Petrini$^c$, M. Porrati$^{a,d,e}$ and
A. Zaffaroni$^a$\footnotemark}
\footnotetext{e-mail: girardello@milano.infn.it, \,m.petrini@ic.ac.uk,\,
massimo.porrati@nyu.edu,\, alberto.zaffaro-\\ ni@cern.ch}
\vskip .1in
(a){\em Theory Division CERN, Ch 1211 Geneva 23, Switzerland}
\vskip .1in
(b){\em Dipartimento di Fisica, Universit\`a di Milano via Celoria 16 I 20133
Milano, Italy\footnotemark}
\footnotetext{Permanent Address}
\vskip .1in
(c){\em Theoretical Physics Group, Blackett Laboratory,
Imperial College, London SW7 2BZ, U.K.}
\vskip .1in
(d){\em Department of Physics, NYU, 4 Washington Pl.,
New York, NY 10003, USA\footnotemark}
\footnotetext{Permanent Address}
\vskip .1in
(e){\em Rockefeller University, New York, NY
10021-6399, USA}
\vskip .1in
\end{center}
\vskip .4in
\begin{center} {\bf ABSTRACT} \end{center}
\begin{quotation}
\noindent
We find new, local, non-supersymmetric conformal field theories obtained by
relevant deformations of the N=4 super Yang Mills theory in the large $N$
limit. We contruct interpolating supergravity solutions that 
naturally represent the flow from the N=4 super Yang Mills UV 
theory to these non-supersymmetric IR fixed points. 
We also study the linearization around the N=4 superconformal point of
N=1 supersymmetric, marginal deformations. We show that they give rise to 
N=1 superconformal fixed points, as expected from field-theoretical arguments.
\end{quotation}
\vfill
CERN-TH/98-323 \\
October 1998
\end{titlepage}
\eject
\noindent
\section{Introduction}
Maldacena's conjecture~\cite{malda} is a new powerful tool to study conformal
field theories in the large $N$, large 't Hooft parameter~\cite{'t}
 regime~\cite{pol,witten}. According to this conjecture, one can describe the
field theory encoding the low-energy dynamics of $N$  branes, present in a
closed superstring theory, in terms of the classical dynamics of the closed
superstrings in the near-horizon geometry generated by the branes. In the
large $N$ limit, one can find dynamical regimes in which the closed superstring
theory is accurately approximated by its low-energy effective supergravity.
The example we will be concerned with is $N$ D3 branes in Type IIB superstring
theory. The open-string sector of this theory is described by an interacting
N=4 super Yang Mills (SYM) theory with gauge group $SU(N)$ plus a free,
decoupled theory with gauge group $U(1)$. The coupling constant of the SYM
theory is $g_{YM}^2=g_S$, and the supergravity approximation holds whenever
$g_SN \gg 1$, $N\rightarrow \infty$.

One natural question one may ask of N=4 SYM is whether it allows for
deformations that either are (super)conformal or flow to an interacting
conformal fixed point. This question cannot be
answered using perturbative field theory,
since the fixed point may occur at large value of the coupling
constant. Supersymmetry gives a handle on the non-perturbative domain; in
ref.~\cite{strassler} it was shown that N=4 SYM theories with gauge group
$SU(N)$, $N>2$ possess a three-parameter family of N=1 supersymmetric
{\em exact} marginal deformations. No non-perturbative results exist about
relevant deformations~\footnotemark.
\footnotetext{A perturbative analysis of mass deformations was carried out
in~\cite{ein}.}

In this paper, we use the Maldacena conjecture to study marginal and relevant
deformations of N=4 SYM. In this approach the question to ask is whether
Type IIB supergravity compactified on $AdS_5\times S^5$ can be deformed to
some other background with isometry $SO(2,4)$. The very existence of such an
isometry guarantees that the open-string sector of Type IIB on such background
is a four-dimensional conformal theory; conformal invariance is the easy part
of our job. The difficult part is to connect the deformation of the
background with an operator of the N=4 SYM theory, and to construct
an $AdS$ analog of the renormalization group flow from the $AdS_5\times S^5$ 
to the new one.

The first problem is addressed in Sections 2 and 3. In Section 2, we recall
the dictionary linking composite operators in SYM to supergravity modes in Type
IIB on $AdS_5\times S^5$. This dictionary allows us to associate supergravity
modes to marginal and relevant perturbations of SYM. Some relevant
perturbations, namely mass deformations for scalars and fermions, can also
be described in the five-dimensional N=8 gauged supergravity obtained
by dimensional reduction of Type IIB on $S^5$. The second part of this Section
is rather
technical, and
may be skipped on a first reading. The dimensionally reduced
supergravity is described in Section 3.  In the five-dimensional theory, one
finds two new solutions with isometry $SO(2,4)$, besides the
maximally supersymmetric one. They are stationary points of the 5-d scalar
potential in which some 5-d scalar fields get a nonzero VEV. In the
SYM/$AdS$ dictionary, these fields correspond precisely to mass perturbations
of the SYM theory. In the 10-d Type IIB theory, the new stationary points of
the 5-d theory are non-supersymmetric compactifications on manifolds with
isometry $SO(5)\times SO(2,4)$ and $SU(3)\times U(1)\times SO(2,4)$,
respectively.

The second problem is answered in Section 4. There, we show that the
new stationary points of the 5-d theory are local conformal
field theories, and we construct interpolating supergravity solutions
joining the new minima with the old one. These interpolating solutions are
most naturally interpreted as describing a renormalization group flow, with
the new theory in the infrared and the N=4 SYM in the ultraviolet. 

In Section 5 we turn to the description of marginal deformations in
the supergravity approximation. Nonperturbative results in field 
theory~\cite{strassler} ensure the existence of exact
marginal deformations of N=4 SYM.
By studying the Killing spinor equations of Type IIB supergravity, we
prove that such marginal deformations can be described in the supergravity
approximation to linear order in the deformation, and we briefly comment on
this result.

\section{The Spectrum of Type IIB on $AdS_5 \times S^5$}
According to the celebrated Maldacena conjecture \cite{malda}, N=4 Yang-Mills
theory has
a dual description as the Type IIB string on $AdS_5\times S^5$. This dual
description can be used when the  superstring is weakly coupled: this
corresponds to the large $N$, large t'Hooft coupling
$g=\sqrt{g_{YM}^2N}$~\footnote{In this paper we indicate the Yang-Mills
coupling constant with
$g_{YM}$ and we use the notation $g$ for the t'Hooft coupling which is relevant
in the large $N$ limit. In the same way, in many formulae, even when non
explicitly noticed, the various coupling constants multiplying composite
operators must be understood as the rescaled coupling constants that remain
finite in the large $N$ limit.} regime of the N=4 Yang-Mills theory. The
perturbative expansions in
$\alpha'$ and in the string coupling constant  correspond to the
double $1/g$ and $1/N$ expansions of the Yang-Mills theory \cite{'t},
respectively.

In the holographic prescription \cite{pol,witten}, supergravity and stringy
modes are associated via boundary values in $AdS_5$ to the set of gauge
invariant composite operators of the conformal field theory. The mass of an
$AdS_5$ state is related to the conformal dimension $E_0$ of the corresponding
operator via a formula involving the Casimirs of the conformal group $SO(2,4)$
\cite{witten,FFZ}. In the case of a scalar, we have
\beq
m^2=E_0(E_0-4).
\eeq{mass}
In the weak coupling regime of string
theory, the stringy states in $AdS_5\times S^5$, or,
equivalently, the CFT operators, split into two disjoint sets. The KK
modes coming from the reduction on $S^5$ fall into short multiplets of N=8
supersymmetry, containing states with maximum spin 2. They correspond to
the algebra
of {\it chiral} operators of the N=4 Yang-Mills theory, with dimensions
protected under renormalization. The stringy states, on the other hand, fall
into long
representations, containing up to spin 4. They have a mass
squared of order $1/\alpha'$ which corresponds, by Eq.~(\ref{mass}), to an
anomalous dimension $\sqrt{g}$. We see that, generically, in the large
$N$ limit
with $g\rightarrow\infty$, the stringy states have infinite anomalous
dimension. They
decouple from the OPE's, and the algebra of {\it chiral} operators closes.

In the large $N$ limit
with $g\rightarrow\infty$, the supergravity on $AdS_5\times S^5$
contains all the information about the N=4 chiral operators.
It is the purpose of this paper to investigate, using the supergravity
description,
the relevant and marginal deformations of N=4 Yang-Mills theory that can be
associated with chiral operators.

The spectrum of KK modes on $AdS_5\times S^5$ was computed in \cite{van}. In
Fig. 1, all the KK scalar states with zero or negative mass square are shown.
According to Eq.~(\ref{mass}), they correspond to marginal or relevant
operators in the CFT. They appear in the harmonic expansion around $S^5$ of the
following fields: the Type IIB dilaton $B$, the complex antisymmetric 
two-form ,$A_{\alpha\beta}$, \footnote{$A_{\alpha\beta}$ 
is a linear combination of the NS-NS and R-R two-forms.} with indices
$\alpha,\beta$ along the five-sphere, and a combination
of the dilation mode of the internal metric, $h^\alpha_\alpha$, and the Type
IIB five-form with indices along $S^5$. The $SU(4)$ representation is indicated
in Fig. 1 near each state.

The KK spectrum was decomposed in representations of
the superconformal algebra in \cite{GunMult}. The multiplets are specified by
an integer $p$ which corresponds to the conformal dimension of the
lowest state.
In Fig. 1, the states in the same multiplet lie on a vertical. Notice that, for
reasons that will soon become obvious, there is no multiplet for $p=1$. The
states with $p=2$ belong to the graviton multiplet in $AdS_5$ and are indicated
by filled circles in Fig. 1. The structure of the generic multiplet, with the
$SU(4)$ quantum numbers and the conformal dimension of the states, can
be found in table 1 of ref.~\cite{GunMult}.

Here we  follow  an approach based on N=4
on-shell superfields \cite{FFZ}, which has the advantage of giving explicitly
both the structure of the multiplets and the CFT operators corresponding to the
supergravity modes.
\begin{figure}
\centerline{\epsfig{figure=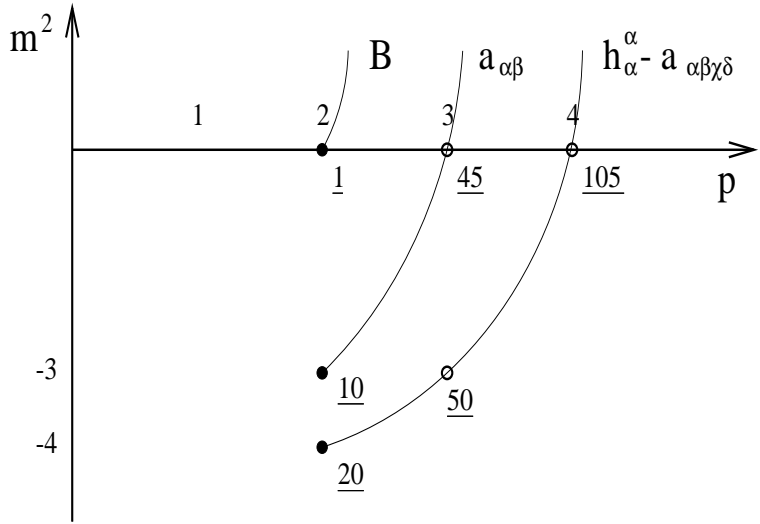,height=7 cm, width=10cm}}
\caption{KK scalar states with zero or negative mass square.
Supersymmetry multiplets
correspond to vertical lines. Filled circles are associated with the states in
the graviton multiplet. The $SU(4)$ representation of each scalar is
indicated, together with the space-time field whose harmonic expansion
gives rise to the particular state.}
\end{figure}
The N=4 chiral multiplets can be obtained by products of the N=4 on-shell
superfield \cite{stelle}.
This superfield, defined in  N=4 superspace, is a Lorentz scalar,
transforming in
the $\underline{6}$ of $SU(4)$ and satisfying some constraints. These
constraints can be found in
\cite{stelle}; they remove all states with spin greater than 1. For our
purpose, we simply need to know that the physical components lie in the first
few terms of the $\theta$ expansion,
\beq
W_{[AB]}=\phi_{[AB]}+\theta_A\lambda_B+\theta_A\theta_B
\sigma^{\mu\nu}_{AB}F^-_{\mu\nu}+ ... + c.c.,\qquad A,B=1,..,4.
\eeq{super}
This is not a chiral superfield, but it satisfies a generalised notion of
chirality when defined in harmonic superspace \cite{howe,FA}. We will also use
the label $i=1,...,6$ instead of the pair of antisymmetric indices $[AB]$.

The spectrum of chiral operators of N=4 Yang-Mills, corresponding to the entire
spectrum of KK states in $AdS_5\times S^5$, is contained in the series of
composite operators $A_p= \Tr W_{\{i_1...i_p\}}-\mbox{traces}$,
obtained by symmetrized
traceless products of the $W_i$ superfield \cite{witten,FFZ}. Due to the trace
on colour space, the first non-trivial
multiplet corresponds to $p=2$. It is the
N=4 multiplet of supercurrents and it is associated with the graviton multiplet
in $AdS_5$. Being conserved, it contains
a  number of states lower than the multiplets with $p\ge 3$; this corresponds
in $AdS_5$ to the fact that the massless multiplet of the graviton sustains a
gauge invariance. By multiplying superfields, it is easy to recover the
structure and the quantum numbers of the states belonging to $p$-th
multiplet, as quoted in table 1 of ref.~\cite{GunMult}.

The superfield $A_p$ has dimension $p$. We can assign dimension $-1/2$ to the
$\theta$'s. It is then obvious that operators with dimension $\le 4$ can be
found only in the very first superfields, namely, $A_2,A_3$ and $A_4$.
The lowest
components of $A_p$ is the scalar $\Tr \phi_{\{i_1}...\phi_{i_p\}}-
\mbox{traces}$, in
the $(0,p,0)$ representation of $SU(4)$ with dimension $p$. For $p\le 4$ we
obtain the states
in Fig. 1, transforming in the $\underline{20},\underline{50}$ and
$\underline{105}$ of $SU(4)$,
with dimensions 2,3 and 4, respectively. They all come from the harmonic
expansion of the same Type IIB field, namely, a linear combination of the
internal
dilation mode and the five-form on $S^5$. In addition to these states, we can
find Lorentz scalars in the $\theta^2$ and $\theta^4$ components of $\Tr W^2$:
these are the states in the $\underline{10}$ and $\underline{1}$ of $SU(4)$ in
Fig. 1, with dimension 3 and 4,
respectively. Finally, there is a scalar in the $\theta^2$ component of
$\Tr W^3$, in the $\underline{45}$ of
$SU(4)$ with dimension 4: it is the last scalar in Fig. 1.
By multiplying superfields, we can write an explicit expression for these
operators,
\bea
(E_0=2,\, \underline{20})\qquad\qquad &\Tr\phi_{\{i}\phi_{j\}}-{\rm traces},\\
(E_0=3,\, \underline{50})\qquad\qquad &\Tr\phi_{\{i}\phi_j\phi_{k\}}-{\rm
traces},\\
(E_0=3,\, \underline{10})\qquad\qquad &\Tr\lambda_A\lambda_B +{\rm cubic}\,
{\rm terms}\, {\rm in}\, \phi,\\
(E_0=4,\, \underline{105})\qquad\qquad
&\Tr\phi_{\{i}\phi_j\phi_k\phi_{t\}}-{\rm traces},\\
(E_0=4,\, \underline{45})\qquad\qquad &\Tr\lambda_A\lambda_B\phi_i +{\rm
quartic}\, {\rm terms}\, {\rm in}\, \phi,\\
(E_0=4,\, \underline{1})\qquad\qquad &{\rm The}\, {\rm N=4}\, {\rm Lagrangian}.
\eea{states}
The cubic and quartic terms in the $\underline{45}$ and $\underline{105}$ come
from the interactions of the N=4 theory. Their explicit expression can be found
by applying the supersymmetry generators to the lowest states of each
multiplet,
or by writing the N=4 theory in N=1 superspace language and solving the
equations of motion for the F and D terms.

The last scalar in Eq.~(\ref{states}) has a special status. Being the highest
component of its multiplet, it preserves N=4 supersymmetry. The corresponding
perturbation of the N=4 theory is simply
a change in the (complexified) coupling constant. It
is an exactly marginal deformation, because the N=4 Yang-Mills theory is
conformal for each value of the coupling. This is consistent
with  the supergravity description, where this operator is the zero mode of the
(complexified) dilaton, which decouples from the Type IIB
equations of motion.

We can study two other marginal deformations of the N=4 theory by
looking at the $\underline{45}$ and \underline{105} operators, and two relevant
deformations corresponding to masses for scalars or fermions (plus additional
cubic terms)
by looking at the $\underline{20}$ and $\underline{10}$ operators.

We can check if the deformations corresponding to the  operators in Fig. 1
preserve some supersymmetry.
This can be easily done by writing the N=4 superfields in N=1 language.
The superfield $W_i$ decomposes into three chiral multiplets $\Sigma_i$,
transforming as a triplet of $SU(3)$ in the decomposition $SU(4)\rightarrow
SU(3)\times U(1)$, and an $SU(3)$-singlet vector-multiplet field strength
$W_\alpha=\bar{D}^2 e^{-V}D_\alpha e^V$,
\beq
W_i=\{\Sigma_i,W_\alpha\}.
\eeq{dec}
The superfield $A_p$, being the product of $p$ fundamental superfields,
decomposes in all possible products of $p$ N=1 superfields $\Sigma_i$ and
$W_\alpha$,
subject to some restrictions coming from total symmetrization and removal
of traces in $A_p$. The complete decomposition for $A_2,A_3$ and $A_4$ can be
found in \cite{FLZ}.

The $\underline{20},\underline{50}$ and $\underline{105}$, which are lowest
components of an $A_p$ superfield, can be written as lowest components of
products of N=1
superfields, $\Sigma_i$ and $\bar\Sigma_i$, and therefore break
supersymmetry completely.

The $\underline{10}$ in $A_2$, on the other hand, decomposes under $SU(3)$ as
$\underline{1}+\underline{3}+\underline{6}$ and corresponds to the components
with dimension 3 in the N=1 superfields
\beq
\underline{1}+\underline{3}+\underline{6}\qquad\qquad \{\Tr W_\alpha
W^\alpha,\Tr \Sigma_iW_\alpha,\Tr \Sigma_i\Sigma_j\}.
\eeq{dieci}
We see that the $\underline{6}$ is the highest component of a 
chiral superfield, and
therefore preserves N=1 supersymmetry. The corresponding deformation is 
a supersymmetric mass term: $\int d^2\theta m_{ij}\Tr \Sigma_i\Sigma_j$. The
$AdS$/CFT
correspondence has little to say about this deformation;
it explicitly breaks conformal invariance and, for a generic $m_{ij}$ flows in
the IR to pure N=1 Yang-Mills which is also not conformal. The other two terms
$\underline{1}$ and $\underline{3}$, being not the highest components of their
N=1 multiplets, break all supersymmetries. We will see however that, despite
the lacking of
powerful supersymmetric methods, we will be able to discuss one of these mass
deformations using N=8 supergravity. We will show in Section 3 and 4 that
the deformation in the $\underline{1}$ (as well as an analogous mode in the
$\underline{20}$ of $SU(4)$) leads in the IR to a
novel N=0 conformal local quantum field theory.

We can repeat the same argument for the $\underline{45}$ in $A_3$ and consider
all the possible products of three N=1 superfields chosen  among $\Sigma_i$ and
$W_\alpha$. 
It is easy to show that in the decomposition $\underline{45}\rightarrow
\underline{10} +\underline{15}+\underline{8}+\underline{6}+
\underline{3}+\underline{\bar 3}$,
the only state that preserves N=1 supersymmetry is $\underline{10}$, being
the highest ($\theta^2$) component of the chiral superfield $\int d\theta^2
\Tr\Sigma_i\Sigma_j\Sigma_k$. We can therefore investigate the following N=1
deformation of the N=4 Yang-Mills theory,
\beq
L_{N=4}+\int d\theta^2 Y_{ijk}\Tr\Sigma_i\Sigma_j\Sigma_k.
\eeq{marg}
The remaining N=1 supersymmetry allows one to derive exact results in the CFT.
It is indeed known \cite{strassler} that there exists a three-dimensional
complex manifold of superconformal N=1 fixed points,
containing N=4 Yang-Mills, as well
as its deformation corresponding to Eq.~(\ref{marg}), and a modification of the
superpotential coupling. The supergravity analysis of this deformation is
the subject of Section 5.

Let us conclude this analysis of the KK spectrum of $AdS_5\times S^5$ in terms
of N=4 Yang-Mills operators by noticing that
the deformations corresponding to {\it chiral} operators by no means exhaust
the class of interesting perturbations of the theory. In the case of relevant
deformations, for example, the mass term for the scalar (the $\underline{20}$),
being traceless, is not the most general one. The diagonal mass term
$\sum_i\Tr\phi_i^2$, is not chiral. It is indeed the simplest example of a
non-chiral operator, since it is the lowest component of
the long multiplet $\Tr W_iW_i$, that contains the Konishi current.
In the case of
marginal deformations, the simplest deformation of N=4 to N=1 corresponds,
in N=1 language, to changing the coefficient of the superpotential, and it is
non-chiral:
\beq
L_{N=4}\rightarrow L_{N=1 DEF}=\int d\theta^2d\bar\theta^2 \Tr
\bar\Sigma_i e^V \Sigma_i+\int d\theta^2{1\over g^2}\Tr W_\alpha W^\alpha +
h\epsilon_{ijk}\Tr\Sigma_i\Sigma_j\Sigma_k + ...
\eeq{enne}
Indeed, the superpotential of N=4 belongs to the
long Konishi multiplet, as the N=1 superfield equation of motion implies
\beq
{\cal W}\equiv \epsilon_{ijk}\Tr\Sigma_i\Sigma_j\Sigma_k=\bar
D^2(\Tr\bar\Sigma_i\Sigma_i).
\eeq{kon}
It is clear that all these non-chiral deformations cannot be easily described
in terms of low energy Type IIB supergravity, because the corresponding $AdS_5$
modes are stringy modes. This will be discussed further in Section 5.

\section{Gauged N=8 Supergravity in Five Dimensions and Relevant Perturbations}
The fact that the low-energy Lagrangian for the states in the graviton
multiplet exists in the form of N=8 gauged supergravity \cite{vantwo} enables
us
to study the deformations corresponding to the $\underline{1},\underline{10}$
and $\underline{20}$ directly from a Lagrangian point of view. These 42 scalars
have a non-trivial potential, which was studied in \cite{gun}. There is a
stationary point
of the potential when all scalar VEVS are zero, with unbroken $SU(4)$
gauge group. This corresponds to the N=4 Yang-Mills theory. We know almost
everything about the deformation in the $\underline{1}$ . It corresponds to
moving along the complex line of fixed points parametrised by the complex
coupling constant. As it must be, on the supergravity side, the
gauged N=8 Lagrangian has a potential which is invariant under
$SU(1,1)$. In other terms, $\underline{1}$ is a flat direction.

The interesting point is that, beside the maximally $SO(6)$-symmetric case,
there are other isolated stationary points
of the N=8 gauged supergravity, corresponding to VEVs
of the $\underline{20}$ and $\underline{10}$. More precisely, two other
stationary points
with metric $AdS_5$ were found in \cite{gun}. Both of them completely break
supersymmetry. In the spirit of the Maldacena $AdS$/CFT correspondence, we are
tempted to interpret these two new $AdS_5$ minima as
corresponding to two N=0 conformal field theories. The fact that they can be
obtained in the N=8 gauged supergravity by giving VEVs to some scalars
can be interpreted as the fact that there is some relevant deformation of N=4
Yang-Mills, which flows in the IR to these novel conformal theories. The CFT
operators associated with the $\underline{10}$ and $\underline{20}$, as
discussed
in the previous Section, are masses for the N=4 Yang-Mills fermions or
scalars. It must be noticed that, since the new minima are not continuously
connected to the maximally symmetric one, the linearization around the
N=4 theory
can not be completely trusted. Higher order corrections, giving rise
to higher dimensional operators in the deformed N=4 theory, must be included.
As a conclusion, we do not know an explicit Lagrangian realization for these
theories, but, as we will show in the next Section, we can prove their
existence as local quantum field theories,
and use supergravity to predict their symmetries.

In \cite{gun} it was suggested that these two N=0 solution correspond to
explicit known compactifications of the type IIB string.

The first stationary point corresponds to
a VEV for the $\underline{20}$ which preserves an $SO(5)$ subgroup of
$SO(6)$. The $AdS$ gauge group is identified with the global symmetry of
the conformal field
theory. According to the general philosophy of the $AdS$/CFT, we expect
that this
CFT, with $SO(5)$ global symmetry, corresponds to some compactification of the
Type IIB string on a manifold with isometry $SO(2,4)\times SO(5)$.
Luckily enough, a manifold with the right properties was
identified in \cite{vNW}, as noticed in \cite{gun}. Topologically, it is
a direct product $AdS_5\times H$, where $H$ is a compact manifold. Metrically,
it cannot be written as a direct product; rather, the $AdS$ metric is
multiplied by a ``warp factor'' depending only on the coordinates
of $H$. We denote such a manifold by $AdS_5\times_W H$. It can
be continuously connected to $S^5$ in the following sense: there exists (at
least) a one-parameter class of manifolds $AdS_5\times_W H(\alpha)$ that solves
the Type IIB equations of motion only for $\alpha=0$  and, say, $\alpha=1$.
At $\alpha=0$ the solution reduces to $AdS_5\times S^5$; at $\alpha=1$
it reduces to $AdS_5\times_W H$~\cite{vNW}. Moreover, the linearization around
$S^5$ shows that we are deforming with a non-trivial
dilation mode of the $S^5$ metric, whose harmonic expansion gives rise,
among other things, to our deformation in the $\underline{20}$, as shown in
Fig. 1.
This strongly suggests that the solution of \cite{vNW} can be identified
with the stationary point found by \cite{gun}.

The second stationary point corresponds to a VEV in the $\underline{10}$, which
preserves an $SU(3)$ subgroup of $SU(4)$. This is the mode
$\underline{1}$, lowest
component of the superfield $W_\alpha W^\alpha$ in Eq.~(\ref{dieci}). As shown
in \cite{gun}, the minimum leaves an unbroken gauge group $SU(3)\times U(1)$,
where the $U(1$) is a combination of the $U(1)$ in the decomposition
$SU(4)\rightarrow SU(3)\times U(1)$ and a $U(1)$ subgroup of $SU(1,1)$. We see
that the $SL(2;Z)$ symmetry of the N=4 theory must play an important role in
the definition of this N=0 theory. Luckily again, a candidate for the
manifold
$H$ exists also in this case. It was found in \cite{R}. It has the same
property as
the previous solution, namely, it is connected to the maximal $S^5$ case by
a one parameter series of manifolds $H(\alpha)$,
with the same properties as before.
By linearising around $S^5$, we identify the deformation with
a non-zero value of the two-form antisymmetric tensor on $S^5$, which gives
rise, after KK expansions, to our mode $\underline{10}$. The $U(1)$
factor among the
symmetries of the solution, which was initially overlooked in \cite{R},
involves a combination of a geometrical $U(1)$ and the $U(1)$ subgroup of
$SU(1,1)$, as noticed in
\cite{gun}, in complete analogy with the supergravity analysis. This again
strongly suggests that the solution of \cite{R} can be identified with the
stationary point found in \cite{gun}.

The explicit parametrisation of the potential and the value of the cosmological
constant at the stationary points will be discussed in the following Section.

Both N=0 conformal theories should belong to a complex line of fixed points
because of the $SU(1,1)$
invariance of the
supergravity minima. In addition to that, supergravity on $AdS_5\times_W H$
should be interpreted as a strong coupling limit, large $N$ expansion of these
theories.
We do not know much about these theories, but the existence of corresponding
Type IIB compactifications allows us to
predict, by analysing the KK excitations, the full spectrum of operators which
have finite dimension in the large $N$, strong coupling expansion of the
theories. In the same limit, the Type IIB equations of motion would reproduce
the Green functions of the theories, according to the holographic prescription
\cite{pol, witten}. The study of the spectrum and of the Green functions in
this particular regime is reduced to the study of a classical supergravity
theory.

\section{The Renormalization Group Flow}
In the previous Section we have found three stationary points of the gauged N=8
supergravity in five dimensions. One, with N=8 supersymmetry, corresponds
to the standard compactification of Type IIB supergravity on the round
five-sphere and is $SO(6)$-symmetric. The other two are non-supersymmetric
and preserve $SO(5)$ and $SU(3)\times U(1)$, respectively. Even though these
stationary points were derived by minimisation of the potential of the
dimensionally reduced theory,
they are most probably true compactifications of Type IIB supergravity
\cite{vNW,R}~\footnotemark.
\footnotetext{This conclusion is also supported by the fact that in
the analogous case of 4-d gauged $SO(8)$ supergravity this is
a {\em theorem}~\cite{dW}: any stationary point of the 4-d superpotential
can be lifted to a solution of 11-d supergravity.}

In the $SO(5)$-symmetric stationary point, some scalars in the $\underline{20}$
of
$SU(4)$ get a nonzero expectation value; while in the
$SU(3)\times U(1)$-symmetric vacuum scalars in the $\underline{10}$
of $SU(4)$ have nonzero expectation values.

In both stationary points the five-dimensional space-time is $AdS_5$,
thus, the 4-d
boundary theory associated with these stationary points is automatically
conformally
invariant. We are not guaranteed {\em a priori} that the boundary theory is
a {local} quantum field theory. A counterexample is the linear-dilaton
background describing the near-horizon geometry of $N$ NS
fivebranes at $g_s\rightarrow 0$~\cite{ABKS}.

By associating a 10-d type IIB background of the form $AdS_5\times_W X$
to these new stationary points, locality follows because
the construction of local operators in the boundary theory proceeds 
by looking at the asymptotic behavior of perturbations in the
interior~\footnotemark.  
\footnotetext{We thank E. Witten for this remark.} 

In this Section, we will prove locality in a different way.
Namely, we will use the UV/IR
relations of $AdS$ dynamics to find an analog to the renormalization
group flow. We will find a solution of N=8 5-d gauged supergravity with two
asymptotic regions, in which the scalar fields depend on the radial $AdS$
coordinate $U$ as follows: in the region near the $AdS$ horizon ($U$
small) the scalars are close to the new stationary point, while as $U$
increases they
roll towards the $SO(6)$-symmetric stationary point. Since $U$, the distance
from the
horizon, is always linearly proportional to the energy scale of the boundary
theory~\cite{PP} one can interpret this solution as describing the RG
evolution from an instable IR fixed point (the new stationary point, non
supersymmetric), towards a stable UV fixed point: N=4 Super Yang Mills.
The UV theory is a local field theory; therefore, the IR fixed point
is also local. Besides helping with locality, our construction also gives a 
function equal to the central charge at the critical points of the scalar
potential, and always increasing along the IR$\rightarrow$UV RG flow.

Both the new stationary points of 5-d $AdS$ supergravity can be obtained by
giving
a nonzero VEV to a single real scalar, that breaks the $SO(6)$ symmetry to
$SO(5)$ or $SU(3)\times U(1)$. Also, in both cases, one can consistently put to
zero all fields except the 5-d metric $g_{IJ}$ and the scalar $\lambda$.
This is possible because all other scalar fields are nonsinglets of the
residual symmetry and must by consequence appear at least quadratically in the
action. Also, with the explicit parametrisation given in \cite{gun}, it
can be checked that these  scalars have no current, so the coupling to
vector fields is also at least quadratic.

The 4-d Poincar\'e-invariant ansatz for the metric is:
\beq
ds^2= e^{2\phi(\rho)}(d\rho^2 + dx^\mu dx_\mu), \;\;\; \mu,\nu=0,..3.
\eeq{m1}
Here $\rho=1/U$, and the $AdS$ background is $e^\phi=U$.
The Lagrangian density of the scalar $\lambda$ can be written as
\beq
L= {1\over 2}mg^{IJ}\partial_I\lambda \partial_J \lambda + V(\lambda),
\eeq{m2}
where $m$ is a nonzero constant.
In our ansatz, the scalar too depends only on $\rho$. The Einstein tensor
$G_{IJ}=R_{IJ}-1/2g_{IJ}R$ has only two independent nonzero components
\beq
G_{\rho\rho}=6{d\phi\over d\rho}{d\phi\over d\rho},\;\;\;
G_{00}=-3{d\phi\over d\rho}{d\phi\over d\rho}
-3{d^2\phi\over d\rho^2}.
\eeq{m3}
Einstein's equations reduce to:
\bea
 G_{\rho\rho} &=& m{d\lambda\over d\rho}{d\lambda\over d\rho}
 - 2e^{2\phi}V(\lambda),\label{m4}
\\
 G_{00} &=& m{d\lambda\over d\rho} {d\lambda\over d\rho}
+ 2e^{2\phi}V(\lambda).
\eea{m5}
The scalar's equation of motion is instead:
\beq
m{d\over d\rho} \left(e^{3\phi}{d\over d\rho}\lambda\right)=e^{5\phi}V'
(\lambda),
\eeq{m6}
(the prime denotes derivative with respect to $\lambda$).

The last  equation is not independent;
rather, it is a linear combination of Einstein's equations.
It is convenient to change variable from $\rho$
to $x$ such that $\exp(-\phi)=d\rho/dx$, and to choose as independent equations
the equations of motion of the scalar and of $g_{\rho\rho}$
\beq
m{d^2\lambda\over dx^2} +4m{d\phi\over dx}{d\lambda\over dx}=V'(\lambda),\;\;\;
6{d\phi\over dx}{d\phi\over dx}= m{d\lambda\over dx}{d\lambda\over dx}
-2 V(\lambda).
\eeq{m7}
By solving with respect to $d\phi/dx$ we find the single equation:
\beq
m{d^2\lambda\over dx^2} \pm{4\over \sqrt{6}}m
\sqrt{m{d\lambda\over dx}{d\lambda\over dx} -2V(\lambda)}{d\lambda\over dx}
=V'(\lambda).
\eeq{m8}
The sign in Eq.~(\ref{m8}) is fixed to $+$ as follows. 1) The equations of
motion~(\ref{m4},\ref{m5},\ref{m6}) imply that $d^2\phi/d^2x\leq 0$.
2) We are looking for a solution that approaches the $SO(6)$ symmetric
$AdS$ solution in the far ``future'' i.e. for $x\rightarrow \infty$.
Moreover we want that increasing $x$ corresponds to increasing energy, i.e.
distance from the $AdS$ horizon; this means
$\phi(x)\rightarrow x/R_2$ when $x\rightarrow \infty$. $R_2$ is the $AdS$
radius. From 1) and 2) it follows that $d\phi/dx\geq 1/R_2>0$, always.

Eq.~(\ref{m8}) has a simple interpretation: it describes the motion of a
particle of mass $m$ in the potential $-V$, subject to a damping with
never-vanishing coefficient $4d\phi/dx$.

In ref.~\cite{gun}, an explicit parametrisation was given for the scalar field
configuration that breaks the $SU(4)$ symmetry of $AdS$ 5-d supergravity to
$SO(5)$ or $SU(3)\times U(1)$. Let us use that parametrisation and analyse
the $SO(5)$ case first.

By calling $\lambda$ the real scalar in the $\underline{20}$ of $SU(4)$ that
gets a nonzero VEV at the $SO(5)$-symmetric minimum, we obtain a Lagrangian
of the form given in Eq.~(\ref{m2}), with $m=45/12$ and a potential~\cite{gun}
\beq
V(\lambda)=-{1\over 32}g^2(15e^{2\lambda}+10e^{-4\lambda}
-e^{-10\lambda}).
\eeq{m9}

It has two stationary points. At the first point, $SO(6)$ symmetric,
$\lambda=0$ and $V=-3g^2/4$. At the other, the symmetry is $SO(5)$,
$\lambda=-(\log 3)/6$, and $V=-3^{5/3}g^2/8$.

We want to prove that there exists an interpolating solution leaving $\lambda=
-(\log 3)/6$ at $x=-\infty$ and stopping at $\lambda=0$ at $x=+\infty$.
This is obvious since that solution describes a particle subject to a
never-vanishing damping moving away from a local maximum of the upside-down
potential ($\lambda=-(\log 3)/6$, $-V=3^{5/3}g^2/8$) and rolling to rest at a
local minimum ($\lambda=0$, $-V=3g^2/4$).
The shape of the upside-down potential, shown in Fig. 2, also shows that the
interpolating solution is generic; namely, that by {\em increasing} $\lambda$
by an arbitrary small amount at $x=-\infty$, one always reaches $\lambda=0$
at $x=+\infty$.
\begin{figure}
\centerline{\epsfig{figure=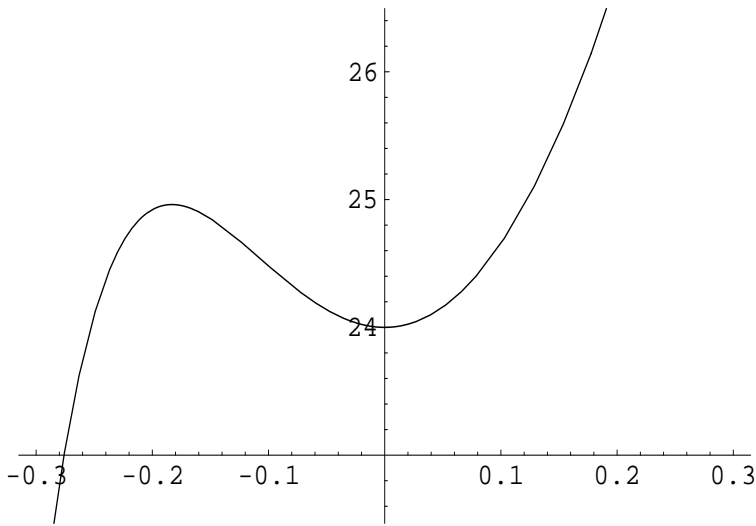,height=7cm, width=10cm}}
\caption{Shape of the $SO(5)$-symmetric upside-down potential $-V(\lambda)$;
units on the coordinate axis are conventional}
\end{figure}

The same argument can be applied verbatim to the case of the
$SU(3)\times U(1)$ symmetric deformation. Denoting by $\lambda$ the
scalar that breaks $SO(6)$ to $SU(3)\times U(1)$ one finds again a Lagrangian
as in Eq.~(\ref{m8}), with a potential~\cite{gun}
\beq
V(\lambda)={3\over 32}g^2 \left[ \cosh(4\lambda)^2 - 
4\cosh(4\lambda) - 5\right]
{}.
\eeq{m11}
This potential is even in $\lambda$. It has three stationary points; that at
$\lambda=0$ is the old $SO(6)$ symmetric one, with $-V=3g^2/4$, while those at
$\cosh(4\lambda)=2$ are $SU(3)\times U(1)$ symmetric, and there $-V=27g^2/32$.
Again, the existence of a solution interpolating between an $SU(3)\times U(1)$
symmetric vacuum and the $\lambda=0$ one is obvious. Also, as in the previous
case, it is generic: any small perturbation such that $\cosh(4\lambda)<2$ at
$x=-\infty$ gives rise to a $\lambda(x)$ that rolls towards $\lambda=0$, and
stops there at $x=+\infty$. The shape of the potential $-V(\lambda)$ is shown
in Fig. 3.
\begin{figure}
\centerline{\epsfig{figure=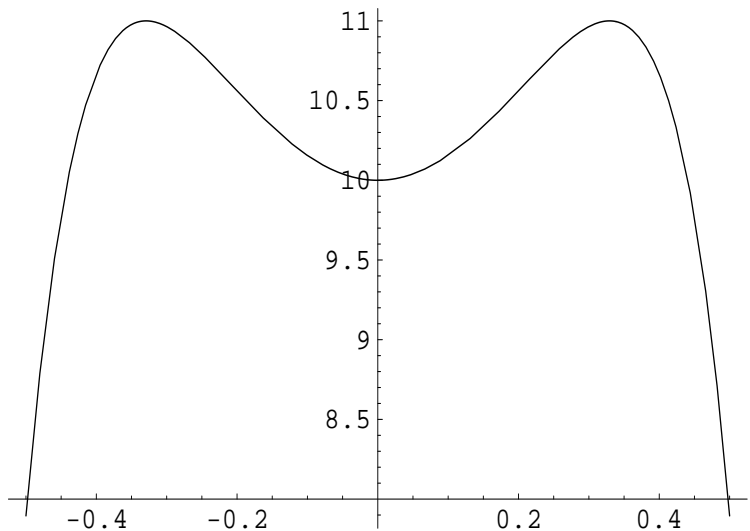,height=7 cm, width=10cm}}
\caption{Shape of the $SU(3)\times U(1)$-symmetric upside-down
potential $-V(\lambda)$}
\end{figure}

The existence of generic interpolating solutions means that N=8 gauged,
5-d supergravity admits solutions on space-times with two asymptotic $AdS$
regions, one near and the other far from the horizon. The holographic
correspondence with 4-d boundary CFTs, and the link that exists between
the distance from the horizon and the energy scale of the boundary theory,
tell us that this solution can be interpreted as a renormalization group flow.
Namely, it corresponds to a flow from an IR non-supersymmetric theory to
N=4 super Yang Mills. The novel IR theories may not have a Lagrangian
formulation, but they are local, as explained at the beginning of this 
Section. Also, our construction of an RG flow explicitly
gives a function, $c(x)$, that equals 
the central charge at the critical  points of the scalar potential, and
that obeys a c-theorem, i.e. that increases along the RG trajectory:
\beq
c(x)= {\rm const\,} (T_{xx})^{-3/2}.
\eeq{ctheo}
At the critical points, the kinetic energy is zero and $T_{\rho\rho}$
equals minus the scalar potential, $V_{crit}$. In ref.~\cite{skend} it was
shown that precisely $V_{crit}^{-3/2}$ is proportional to 
the central charge.
     
In both cases, we found that the new theories are UV instable: in the UV
they flow back to N=4. Conversely, this means that an {\em appropriately
chosen} small perturbation of N=4 will flow in the IR to one of our
novel theories. To identify completely the perturbation goes beyond the
possibilities of today's $AdS$/CFT techniques. This identification would
require
a complete control over all non-renormalizable operators present in N=4.
We can nevertheless expand the perturbation in a power
series of the perturbation
parameter, $\epsilon$, and identify the first term in the series.

In the first example we examined, we gave a nonzero expectation value to
a supergravity scalar in the $\underline{20}$ of $SU(4)$. This scalar has
$AdS$ mass square equal to $-4$. Using the
correspondence between supergravity fields and composite operators in N=4
established in the previous Sections, we identify this perturbation with
a composite operator of N=4 super Yang Mills of dimension 2:
a mass term for the scalars $\phi_i$,
symmetric and traceless in the index $i$.
Thus the perturbation breaking $SO(6)$ to $SO(5)$ is:
\beq
{\cal O}=\epsilon \left(\sum_{i=1}^5\Tr \phi_i^2 -5\Tr \phi_6^2\right) +
O(\epsilon^2).
\eeq{m12}
The trace is taken over the gauge-group indices;
the $O(\epsilon^2)$ terms are higher-dimension operators that, among other
things, stabilise the runaway direction $\phi_6$.

In the second example, we gave an expectation value to a supergravity scalar
in the $\underline{10}$ of
$SU(4)$, with $AdS$ mass square equal to $-3$.
In the N=4 CFT it roughly corresponds
to a mass term for the fermion $\lambda^4$ (the N=1 gaugino).
Thus, the perturbation breaking $SO(6)$ to $SU(3)\times U(1)$ is
\beq
{\cal O}=\epsilon \left (\Tr \lambda^4 \lambda^4 +
\mbox{cubic terms in}\, \phi_i\right ) + O(\epsilon^2) + h.c..
\eeq{m13}
\vskip .1in
\noindent
{\bf Addendum}
\vskip .1in
\noindent
A few days after this paper was posted on the web, a paper appeared~\cite{DZ} 
that closely parallels the results of this Section. In that paper, the 
spectrum of our supergravity solutions was also computed, with the result that
while the 
$SU(3)\times U(1)$ stationary point is stable, the $SO(5)$ is not. Namely, a
scalar in the 14 of $SO(5)$ has a negative mass square exceeding the 
Breitenlohner-Freedman bound~\cite{BF}. The corresponding field in the 
boundary conformal field theory would have then a complex conformal weight.

This result is puzzling because one may think that the infrared fixed point of
a positive, local field theory (N=4 SYM) should not exhibit such pathology.
We do not have a definite answer to this puzzle; we may just notice that
the $SO(5)$ point was obtained in the first place by a somewhat peculiar
perturbation, Eq.~(\ref{m12}), which is tachionic at zero scalar VEVs.
This tachion signals that under this perturbation the SYM scalars must get a 
nonzero VEV, and the configuration with $N$ coincident 3-branes is unstable 
and must break apart.
The instability of the $SO(5)$-symmetric point probably means that its
vacuum state has a finite lifetime. This conjecture is supported by the 
following observation.

Let us call $\phi$ all supergravity fields on $AdS_5$, and 
$Z[\phi_0]$ the supergravity partition function on $AdS_5$, computed with 
the boundary condition that $\phi=\phi_0$ at infinity. If $\phi_0$ is a 
stationary point of the supergravity scalar potential, the $AdS$/CFT 
correspondence states that 
\beq
Z[\phi_0]=\langle 0| 0 \rangle_{CFT},
\eeq{mm1}
where $|0\rangle $ is the vacuum state of the CFT defined by the 
stationary point of the scalar potential.
At leading order in the $1/N$ expansion, $Z[\phi_0]=\exp\{-S[\phi_0]\}$, where
$S[\phi_0]$ is the classical supergravity action. This action is real, but
to next order in the expansion one gets 
$Z[\phi_0]=\exp\{-S[\phi_0] -1/2{\rm Str}\,\log S''[\phi_0]\}$, with 
$S''[\phi_0]$ the matrix of the quadratic fluctuations around 
$\phi_0$. The supertrace is complex when there exist fluctuations 
that do not satisfy the Breitenlohner-Freedman bound. Its imaginary part
is {\em nonvanishing} in the large $N$ limit; therefore, the 
lifetime of the vacuum is finite. Thanks to Eq.~(\ref{mm1}), 
the vacuum  decay rate per unit volume is:
$\Gamma = (2VT)^{-1}\Im{\rm Str}\,\log S''[\phi_0]$, where $VT$ is the 4-d 
space-time volume~\footnotemark.
\footnotetext{A somewhat related analysis of conditions for 
the stability of the large $N$ supergravity approximation of 
non-supersymmetric theories was performed in~\cite{BR}.} 
\section{Marginal Deformations of N=4 Super Yang Mills in the Supergravity
Limit}
It is known that there exists a manifold of N=1 fixed points that contains the
N=4 Yang-Mills theory \cite{strassler}. The corresponding theories can be
described in N=1 language as containing the same fields as N=4 but with a
superpotential
\beq
{\cal W}= h\epsilon_{ijk}\Tr\Sigma_i\Sigma_j\Sigma_k +
Y_{ijk}\Tr\Sigma_i\Sigma_j\Sigma_k,
\eeq{leigh}
where $Y_{ijk}$ is a generic symmetric tensor of $SU(3)$, or, in other terms,
an element of the $\underline{10}$ of $SU(3)$. The N=4 theory is recovered for
$Y_{ijk}=0$ and $h=g$. There is a particular relation between $g,h,Y_{ijk}$ for
which
the theory is superconformal. The reason is very simple \cite{strassler}.
The theory is conformal if the anomalous
dimension matrix $\gamma_i^j$ for the matter fields $\Sigma_i$
vanishes. The reason is that non-perturbative exact formulae \cite{shif}
relate  the N=1 gauge beta function to the anomalous dimensions of the matter
fields. When the N=1 non-renormalization theorems are used, the vanishing of
$\gamma_i^j$ is enough to guarantee the vanishing of the beta
functions for all the couplings of the theory. In our case, with 24 real
parameters, the 9
conditions
\beq
\gamma_i^j(g,h,Y_{ijk})=0,
\eeq{anom}
combined with the modding  by $SU(3)$ and the $U(1)$ R-symmetry, yield a
three-complex dimensional manifold of fixed points.

If the N=4 Yang-Mills theory can be embedded in this larger manifold of N=1
fixed points, we should be able to find a continuous family of solutions of the
Type IIB string theory associated with backgrounds of the type $AdS_5\times_W
H(g,h,Y_{ijk})$, continuously connected to $AdS_5\times S^5$ \footnote{Notice
that, unlike the discussion in Section 3, we are now looking for backgrounds
that are Type IIB solution for each value of the parameters and not only for
particular ones.}. We will take a perturbative point of view: if we can
identify $h$ and $Y_{ijk}$ with KK modes in the spectrum of
$AdS_5\times S^5$,
we can deform the symmetric solution by turning on the corresponding $S^5$
harmonic, and check order by order in a perturbative expansion in $h$ and
$Y_{ijk}$ whether or not the new background is a solution of the Type IIB
equations of motion\footnote{We thank O. Aharony
for pointing out a mistake in an earlier version of this manuscript, that
changed the conclusions of this Section.}.

We discussed the $AdS_5$ interpretation of both terms in~(\ref{leigh}) in
Section 1. The conclusion was that, while the deformation in the
$\underline{10}$ can be identified with a particular KK mode (part of the
$\underline{45}$ in Fig. 1) and can be therefore studied in the supergravity
approximation, the
coupling $h$ of the N=4 superpotential must be associated with a string state,
which we are not able to study within the supergravity approximation.
At first order in the deformation this is not a problem, since 
we can consistently study the case $g=h$ in the supergravity
limit, and $g-h$ is quadratic in the deformation. 
Next, we turn to the explicit supergravity calculation.

As already explained in Section 1, the marginal deformation $Y_{ijk}$ of N=4
Super Yang-Mills theory can be identified with part of the KK scalar mode in
the $\underline{45}$ of $SO(6)$. More precisely, this scalar corresponds
to the second
two-form harmonic
$Y_{[\alpha\beta]}^I$ ($k=2$ in the language of ref.~\cite{van})
in the expansion of the antisymmetric two-form, $A_{\alpha\beta}$, with
components along the five-sphere~\cite{van}
\beq
A_{\alpha\beta}=\sum a^I(x)Y_{[\alpha\beta]}^I(\theta_{\alpha}).
\eeq{nnn}
Here $x$ is the $AdS$ coordinate and $\theta_{\alpha}$ is a set of
angular coordinates on $S^5$. $I$ is an index labelling the $SU(4)$
representation of the harmonic $Y$; in this case, $I$ labels the
$\underline{45}$ of $SU(4)$.
We can then try to construct a new set of supergravity solutions by
turning on this mode. The ansatz for our solution is, to linear
order in the deformation,
\bea
g_{MN} &=&\dot{g}_{MN}\nonumber\\
A_{MN} &=& \left\{
       { a^I Y_{[\alpha\beta]}^I  \qquad\mbox{for}\,  M,N=\alpha,\beta=5,..,9;}
       \atop
       { 0 \qquad \qquad  \qquad \qquad \mbox{otherwise}.\qquad}
       \right. \nonumber\\
F_{MNRST}&=& \left\{
               {\epsilon_{\mu\nu\rho\sigma\tau},\qquad \mu,\nu=0,..,4,}
               \atop
               {\epsilon_{\alpha\beta\gamma\delta\epsilon},\qquad \alpha,\beta
=5,..,9,}
               \right. \nonumber\\
B&=&\mbox{cost}.
\eea{bbbb}
Here $\dot g_{MN}$ is the metric on $AdS_5\times S^5$ and all fermion
fields are set to zero.

We know from \cite{van} that, at the linearised
level, the Type IIB equations of motion reduce to:
\beq
D_\mu D^\mu a^I(x)=0.
\eeq{lap}
A field with zero mass in $AdS_5$ corresponds to a marginal operator in the
CFT, according to Eq.~(\ref{mass}).
Since we are looking for supergravity solutions corresponding to conformal
quantum field theory, we do not want the $AdS$ geometry of the five
dimensional space-time to be modified.
Therefore we take the coefficient $a^I$ independent
of the $AdS$ coordinates $x$. This is obviously a solution of the Type IIB
equations of motion.

To check that this deformation is indeed a solution of Type IIB compactified
on $AdS_5$ times a continuous deformation of $S^5$, we should in
principle verify that it satisfies, order by order in the perturbative
parameter $a^I$, the type IIB equations of motion.
Moreover, this new background must be supersymmetric.
On the field theory side of the $AdS$/CFT correspondence, indeed, we
want a family of N=1 superconformal theories; thus,
the supergravity solution must have N=2 supersymmetry.
In such a background, the supersymmetry shift
of the fermionic fields must be zero. The general supersymmetry
transformations for the fermions are given in~\cite{schwarz}; to linear
order in the deformation they read:
\bea
&0& =\delta \lambda=-\frac{i}{24} \Gamma^{MNP}
G_{MNP}\dot\epsilon   \label{33}\\
&0&= \delta \psi_M=
                \hat D_M\epsilon^{(1)} +\frac{1}{96}
                \left(\Gamma_M\null^{NPQ}
                G_{NPQ}-
                9\Gamma^{NP}G_{MNP}
                \right)\dot\epsilon^c,
\eea{bbbc}
where $G_{MNP}=3\partial_{[M}A_{NP]}$, $\hat D_M$ is the $AdS_5\times
S^5$ covariant derivative and $\epsilon$
is a ten dimensional left-handed spinor. We also denoted by $\epsilon^c$ the
charge-conjugated spinor. $\dot{\epsilon}$ is the zero-order spinor that
satisfies
the above equation for the maximally symmetric background of
ref.~\cite{van}: $\hat
D_M\dot\epsilon=0$, while $\epsilon^{(1)}$ denotes the first order
correction in the deformation.

We will use the following decomposition for the ten-dimensional
$\Gamma$ matrices
\beq
\Gamma^\mu=\gamma^\mu\otimes {\bf 1}_4\otimes \sigma^1,\,\,
\Gamma^{\alpha}={\bf 1}_4\otimes
\tau^\alpha\otimes (-\sigma^2),
\eeq{gamma}
where $\gamma^\mu$ and $\tau^\alpha$ are two sets of five-dimensional gamma
matrices for the space-time and internal dimensions, respectively.
The reduction to five dimensions is performed by expanding in
harmonics also the supersymmetry spinor, $\epsilon$,
\beq
\epsilon(x,\theta)=\sum \epsilon^I(x)Y_a^I(\theta),
\eeq{epm}
where $a=1,...,4$ denotes a spinorial index on $S^5$. $\dot{\epsilon}$ simply
coincides with the first harmonic of the expansion~(\ref{epm}), which
corresponds
to the representation $\underline{\bar{4}}$ of $SU(4)$: these are the four
complex spinor parameters of N=8 supersymmetry. 

We must satisfy the equations:
\bea
&0& =\delta \lambda=-\frac{i}{24} \Gamma^{\alpha\beta\gamma}\dot{\epsilon}
G_{\alpha\beta\gamma}\label{BB}\\
&0&= \delta \psi_M= \left\{
              {   \hat{D}_{\mu} \epsilon^{(1)} +\frac{1}{96}
                \Gamma_{\mu}\null^{\alpha\beta\gamma}
 G_{\alpha\beta\gamma}\dot{\epsilon}^c\qquad \qquad \qquad}\atop
              {   \hat{D}_{\alpha}\epsilon^{(1)} +\frac{1}{96}
                \left(\Gamma_{\alpha}\null^{\beta\gamma\delta}
                G_{\beta\gamma\delta}-
                9\Gamma^{\beta\gamma}G_{\alpha\beta\gamma}
                \right)\dot{\epsilon}^c  .}
                \right.
\eea{BB1}
Both $G_{\alpha\beta\gamma}$ and $\dot\epsilon$ are known functions on $S^5$
transforming in the $\underline{45}$ and $\underline{\bar 4}$ of $SU(4)$,
respectively. The various products of these functions that appear in
Eq.~(\ref{BB}) can be decomposed into sums of harmonics:
\bea
&0 =\delta\lambda_a =\sum Y^I_a,\qquad &I\in
\underline{45}\times\underline{\bar 4}\rightarrow
 \underline{140}+\underline{20}+\underline{20}'\nonumber\\
&0 =\delta\psi_{a\mu} = \hat D_\mu\epsilon^{(1)} +\Gamma_\mu\sum Y^I_a,\qquad
&I\in \underline{45}\times \underline{4}\rightarrow\underline{84}+
\underline{60}+\underline{36}\nonumber\\
&0 = \delta\psi_{a\alpha} = \hat D_\alpha\epsilon^{(1)} +\sum
Y^I_{a\alpha},\qquad &I\in \underline{45}\times \underline{4}
\rightarrow\underline{84}+\underline{60}+\underline{36}.
\eea{MM}

Not all the representations indicated in Eq.~(\ref{MM}) actually appear. A
generic harmonic $Y^I_i$ of $S^5=SO(6)/SO(5)$ is specified by giving the
representation $I$ under the isometry group of the five-sphere, $SO(6)$,
and the representation $i$ under the local Lorentz group of the sphere,
$SO(5)$. In our
case, the gaugino transforms in the $4$ of $SO(5)$ and the gravitino in the
$4+16$ of  $SO(5)$. As a general rule \cite{salam}, the representation $i$ must
appear in the decomposition of the representation $I$ of $SO(6)$ under $SO(5)$.
In Fig. 4, where the product of the relevant representations is expressed in
terms of $SU(4)$ Young tableaux, the representations containing the $4$ or $16$
of $SO(5)$ are explicitly indicated. We see that, for example, $\underline{84}$
and $\underline{20}$ do not contribute to the right
hand side of Eq.~(\ref{MM}).

\begin{figure}
\centerline{\epsfig{figure=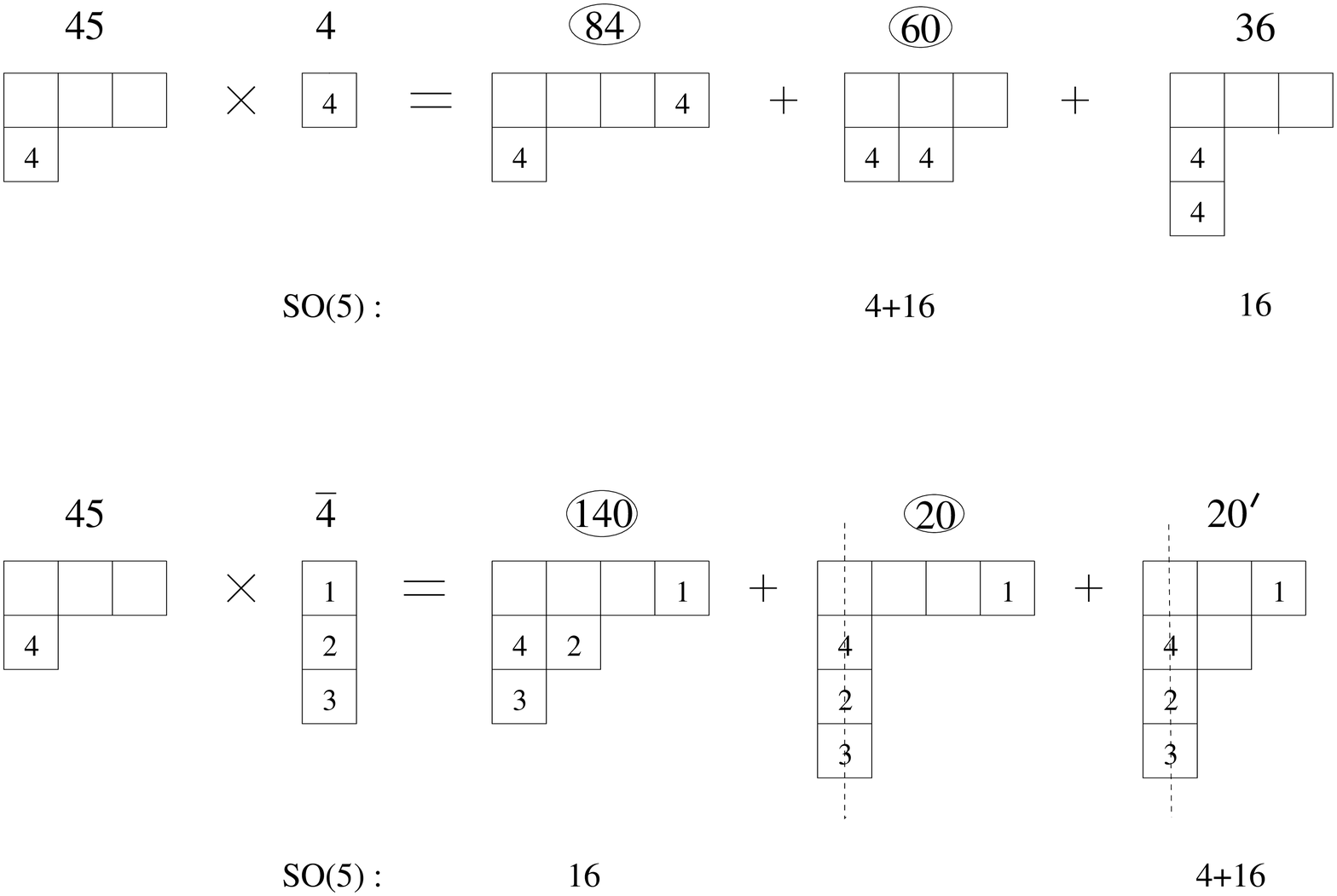,height=7 cm, width=10cm}}
\caption{Products of harmonics in the fermion variations are decomposed
into $SU(4)$ representations. Representations that contain the $4$ or $16$
of $SO(5)$ and can, therefore,
contribute to the expansion into harmonics of the
fermion shifts are indicated. The reduction to
$SU(3)$ is obtained by splitting the indices of the $\underline{4}$ into
$i=1,2,3$ and $4$, and by putting the index $4$ in all the relevant boxes. The
representations that contains a $\underline{10}$ of $SU(3)$ are encircled.}
\end{figure}

Eqs.~(\ref{MM}) can be satisfied to linear order in the deformation
only if
\beq
G_{\alpha\beta\gamma}\Gamma^{\alpha\beta\gamma}\dot{\epsilon}=0.
\eeq{hh}

This equation is simply the gaugino variation. 
It is satisfied for the following reason. 
Only the harmonics transforming in the $4$ of
$SO(5)$ contribute to the equation, thus we see from Fig. 4 that 
only $\underline{20}'$ may contribute. Also, we are not interested in
satisfying these equations for all the 45 complex deformations in
Eq.~(\ref{nnn}), but only for those that transform as $\underline{10}$ under
$SU(3)\subset SU(4)$. Moreover, we are  interested in preserving only one
supersymmetry out of the initial 4: the $SU(3)$ singlet in the decomposition
$\underline{4}\rightarrow \underline{1}+\underline{3}$ under
$SU(3)\subset SU(4)$. Therefore,
we have to consider only the $SU(3)$-terms $\underline{10}\times \underline{1}$
in the products $\underline{45}\times\underline{\bar 4}$.
The representations that contain a $\underline{10}$ are encircled in Fig. 4. 
We see that Eq.~(\ref{hh}) can be satisfied because $\underline{20}'$ 
does not contain the $\underline{10}$ of $SU(3)$. 
Notice that we must choose $\dot{\epsilon}$ in the $\underline{\bar 4}$; 
if we choose it in the $\underline{4}$ Eq.~(\ref{BB}) has no solution. 

To cancel the gravitino shifts in Eq.~(\ref{BB1}) we first recall that by 
definition:
\beq
\left(D_M +{1\over 2}ie\Gamma \Gamma_M\right)\dot\epsilon \equiv 
\hat{D}_M \dot\epsilon=0.
\eeq{kill}
Here $\Gamma=\Gamma_5...\Gamma_9$, and $e$ determines the curvature
of $S^5$. 
From the first of Eqs.~(\ref{BB1}) we find
\beq
\epsilon^{(1)}=-{1\over 96}ie^{-1}\Gamma \Gamma^{\alpha\beta\gamma}
G_{\alpha\beta\gamma}\dot\epsilon^c.
\eeq{e'}
We must check that this expression cancels the gravitino shifts with indices
in $S^5$, given by the second of Eqs.~(\ref{BB1}).
Setting to zero the gamma trace of $\delta\psi_\alpha$ we find: 
\beq
\Gamma^\alpha \delta\psi_\alpha =\left(
\Gamma^\alpha D_\alpha -{9\over 2}ie\Gamma\right)
\epsilon^{(1)}=0.
\eeq{gtr}
This equation holds because the only harmonic contributing to 
$\delta\psi_\alpha$ is the $\underline{60}$, as shown in Fig. 4. In
ref.~\cite{van}, it was shown that the equation satisfied by this spinor 
harmonic ($\Xi^{2,-}$ in the notations of~\cite{van}) is, precisely, 
Eq.~(\ref{gtr}).
The divergence $D^\alpha\delta\psi_\alpha$ gives another constraint on
$\epsilon^{(1)}$. Using the identity
$D^\alpha D_\alpha =(\Gamma^\alpha D_\alpha)^2 -5e^2$, and the 
equation of motion of $G_{\alpha\beta\gamma}$~\cite{schwarz}:
\beq
D^\delta G_{\delta\alpha\beta}=
-{2\over 3}i\epsilon_{\alpha\beta\gamma\delta\epsilon}
G^{\gamma\delta\epsilon},
\eeq{eomg}   
we find, after some elementary gamma-matrix algebra:
\beq
D^\alpha\psi_\alpha=\left[(\Gamma^\alpha D_\alpha)^2 +
{3\over 2}ie\Gamma \Gamma^\alpha D_\alpha +27e^2\right]\epsilon^{(1)}=0.
\eeq{dvg}
This equation is satisfied whenever Eq.~(\ref{gtr}) is.

To prove that $\epsilon^{(1)}$ cancels all gravitino shifts we must show that
$\delta\psi_\alpha$ contains no transverse (i.e. divergenceless and
gamma-transverse) term. This follows again from a result of ref.~\cite{van}:
no transverse harmonic belongs to the $\underline{60}$ of $SU(4)$.

A second, direct proof is obtained by projecting $D_\alpha\epsilon^{(1)}$ on 
a complete basis of transverse vector-spinor harmonics, 
$\Xi_\alpha^{I_T}$ in the notations of~\cite{van}. By definition
$\Xi_\alpha^{I_T}$ obeys $D^\alpha \Xi_\alpha^{I_T}= \Gamma^\alpha 
\Xi_\alpha^{I_T}=0$. Using the Bianchi identity of
$G_{\alpha\beta\gamma}$, its equation of motion, and elementary gamma-matrix
manipulations we find, thanks to the transversality of $\Xi_\alpha^{I_T}$:
\beq
0=\int_{S^5}d^5x \bar{\Xi}^{\alpha\, I_T}(x) D_\alpha\epsilon^{(1)} =
-{1\over 32}ie^{-1}\int_{S^5}d^5x \bar{\Xi}^{\alpha\, I_T}(x)\left(
\Gamma^\delta D_\delta -{7\over 2}ie \Gamma\right)
G_{\alpha\beta\gamma}\Gamma^{\beta\gamma}\dot\epsilon^c.
\eeq{tt}
Integrating by parts, we see that this 
equation means that the only term in the gravitino shift that can contain
a transverse part, 
$G_{\alpha\beta\gamma}\Gamma^{\beta\gamma}\dot\epsilon^c$, is at most 
proportional to the harmonic $\Xi_\alpha^{I_T}$ obeying 
$(\Gamma^\delta D_\delta -{7\over 2}ie \Gamma)\Xi_\alpha^{I_T}=0$.
Since this harmonic is the product of a $\underline{15}$ with a 
$\underline{4}$, it has no component in the $\underline{60}$ of 
$SU(4)$~\cite{van}; therefore, 
$\delta\psi_\alpha$ has no transverse component at all.
 
We have now completed the proof that there exist a linearized deformation
of the $AdS_5\times S^5$ background with $SO(2,4)$ isometry 
and preserving N=2 supersymmetry. 

To linear order, all deformations in the $\underline{10}$ of $SU(3)$, 
$Y_{ijk}$, preserve
N=1 superconformal invariance. We do expect to see the equivalent of 
Eq.~(\ref{anom}) starting to second order in $Y_{ijk}$, similarly to field 
theory. In our case, the role of Eq.~(\ref{anom}) is played by the
type IIB equations of motion, and, in particular, by the Einstein equations. 

As a concluding remark, let us point out that the existence of superconformal
deformations of type IIB on $AdS_5\times S^5$ is far from trivial, 
even to linear order. When seen from the viewpoint of the type IIB theory, 
this result 
implies that on the $AdS$ background an N=8 supersymmetric solution 
can be continuously deformed to solutions with lower supersymmetry. 
This is not possible on a Minkowsky background~\cite{BaDi}. The use of
marginal deformations of type IIB as a means to obtain supersymmetry-changing 
transitions was also put forward in~\cite{st98}.    
\vskip .2in
\noindent
{\bf Acknowledgements}\vskip .1in
\noindent
We would like to thank O. Aharony, D. Anselmi, S. Ferrara, D.Z. Freedman,
Y. Oz, E. Rabinovici and E. Witten for 
useful discussions and comments. 
L. G. would like to thank the Departments of Physics at
NYU and the Rockefeller University for their kind hospitality.  L. G. and
M. Petrini are partially supported by INFN and MURST, and
by the European Commission TMR program ERBFMRX-CT96-0045,
wherein L.G is associated to the University Torino,
and M. Petrini to the Imperial College, London. M. Porrati
is supported in part by NSF grant no. PHY-9722083.

\end{document}